\PassOptionsToPackage{sort&compress}{natbib}
\documentclass{jais}       
\journal{JAIS-ID}
\vol{xx}
\received{xx January 2025}
\published{xx March 2025}
\usepackage{amsmath}
\usepackage{amssymb}
\usepackage{float}
\usepackage[hidelinks]{hyperref}

\def\be{\begin{equation}}
\def\ee{\end{equation}}
\def\bea{\begin{eqnarray}}
\def\eea{\end{eqnarray}}

\renewcommand{\thefootnote}{\fnsymbol{footnote}}

\begin{document}
\title{Sealed, Lightweight gRPC-Based Muon Telescopes for\\ Muography Applications}
\author{
D.~Ahmadi,\auno{1}\footnotemark\
M.~Tytgat,\auno{1}
H.~Basiri,\auno{2}
A.~Biolchini,\auno{2}
E.~Cortina~Gil,\auno{2}
P.~Demin,\auno{2}
R.~Dey,\auno{3}
A.~Giammanco,\auno{2}
S.~Ikram,\auno{2}
R.~Karnam,\auno{3}
V.~K.~S.~Kashyap,\auno{3,4}
B.~Mohanty,\auno{3,4}
and G.~Nyitrai\auno{2}
}

\address{
\begin{flushleft}
$^1$Inter-University Institute for High Energies (IIHE), Vrije Universiteit Brussel, Pleinlaan 2, 1050 Brussels, Belgium\\
$^2$Centre for Cosmology, Particle Physics and Phenomenology (CP3), UCLouvain, Chemin du Cyclotron 2, 1348 Louvain-la-Neuve, Belgium\\
$^3$Centre for Medical and Radiation Physics, National Institute of Science Education and Research Bhubaneswar, An OCC of HBNI, Jatni, Khurda 752050, Odisha, India\\
$^4$School of Physical Sciences, National Institute of Science Education and Research Bhubaneswar, An OCC of HBNI, Jatni, Khurda 752050, Odisha, India
\end{flushleft}
}

\begin{abstract}

\begin{flushleft}
Portable muon detection systems are increasingly in demand for muography applications in environments requiring flexible installation and transportability. We present the ongoing development and implementation of portable muon tracking systems based on glass Resistive Plate Chambers (gRPCs). The work addresses the fabrication and assembly of compact, position-sensitive RPC detectors, along with their performance when operated with a standard gas mixture and coupled to external front-end electronics and a scalable multi-channel data-acquisition system. The system design prioritizes modular construction, mechanical robustness, operational safety, autonomous functionality, and cost-effectiveness.

Measurements with atmospheric muons demonstrated stable detector performance and reliable tracking capability, confirming the suitability of this architecture for muographic measurements. We conclude with an outlook on further developments of the detection system.

\par\vspace{8pt}

\par\vspace{4pt}
\end{flushleft}
\end{abstract}
\maketitle
\renewcommand{\thefootnote}{\fnsymbol{footnote}}
\footnotetext{Corresponding author: donya.ahmadi@vub.be}

\section{Introduction}\label{sec:intro}

\subsection{Muography with cosmic-ray muons}

Cosmic-ray muons are produced in the decay of secondary particles created by primary cosmic-ray interactions in the upper atmosphere, and reach sea level with a mean energy of a few GeV. Owing to their high penetrating power, they can be used to probe the internal structure of large and dense objects. This imaging technique, known as muography, is now applied in a wide range of fields, from Earth sciences to industry and security~\cite{Tanaka2023,Bonechi2020,Bonomi2020}.

Two main approaches are used. Absorption (or transmission) muography measures the flux of muons surviving after crossing an object. The transmitted fraction is related to the density integrated along the muon path, i.e.\ the opacity, which makes the method well suited to large and thick targets. It provides two-dimensional density maps, or three-dimensional reconstructions when several viewpoints are combined~\cite{Procureur2023}, but requires long exposure times to accumulate sufficient statistics. Scattering muography measures the angular deflection of muons due to multiple Coulomb scattering inside the object. Since the scattering angle depends on the density and atomic number $Z$ of the material, this method allows material discrimination and three-dimensional imaging of small and medium-sized targets relatively quickly. It requires, however, tracking detectors on both sides of the target.

Muography is passive and non-destructive and relies on a natural radiation source; its main limitation is the low muon rate, which sets the statistical precision and the required exposure time. Imaging with cosmic-ray muons has found application in a steadily expanding range of disciplines~\cite{IAEA2022}, including archaeological investigations~\cite{Procureur2023Khufu, Olah2022}, the monitoring of volcanic edifices~\cite{DErrico2022}, geoscientific surveys~\cite{Lechmann2021}, monitoring of CO2 geological storage~\cite{BASIRI2026104683}, the non-invasive characterization of nuclear waste~\cite{Weekes2021}, border security and cargo inspection~\cite{Barnes2023}, and non-destructive testing of industrial installations~\cite{MartinezRuizdelArbol2019}.

In practice, field logistics are often a limiting factor, especially in underground, confined or hard-to-reach locations with restricted space, access and power. Portable detectors could potentially overcome these constraints, as they can be transported and deployed close to the object of interest~\cite{Andringa2024,Balazs2024}, thereby extending the use of muography to sites where large and fixed installations cannot be operated.

\begin{figure*}[t]
\centering
\begin{minipage}[b]{0.31\textwidth}
\centering
\includegraphics[width=\linewidth,height=4.5cm,keepaspectratio]{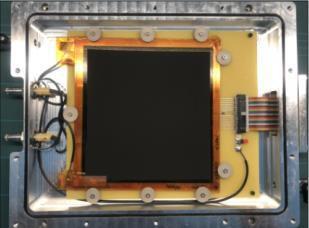}
\end{minipage}\hfill
\begin{minipage}[b]{0.31\textwidth}
\centering
\includegraphics[width=\linewidth,height=4.5cm,keepaspectratio]{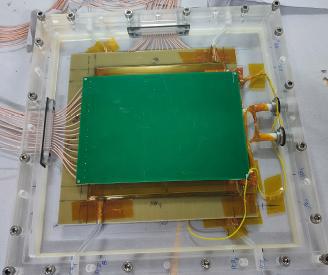}
\end{minipage}\hfill
\begin{minipage}[b]{0.35\textwidth}
\centering
\includegraphics[width=\linewidth,height=4.5cm,keepaspectratio]{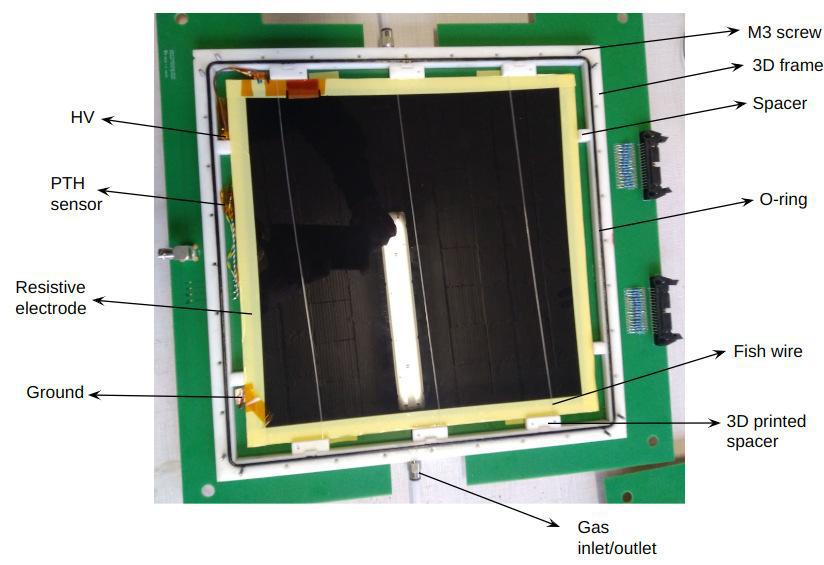}
\end{minipage}
\caption{Internal view of the three gRPC prototypes.}
\label{fig:layouts}
\end{figure*}

\subsection{Resistive plate chambers for muography}
Resistive plate chambers (RPCs) are gaseous detectors consisting of two parallel resistive electrodes separated by a gas gap of a few millimeters. They combine good spatial resolution, typically between 300~$\mu$m and 1~mm depending on the readout segmentation, with time resolutions ranging from about 100~ps to 1~ns, determined mainly by the gap thickness and number of gaps. Their active area can be scaled from a few cm$^2$ to m$^2$ without
compromising performance in terms of e.g.
position resolution or homogeneity. In addition, RPCs are low-cost, mechanically robust and straightforward to construct, and their signals can be read out with standard front-end electronics and data-acquisition (DAQ) systems. These properties make them well suited for portable tracking detectors required in muography.

Gaseous detectors have already been used in several muography campaigns, for example, 1~m$^2$ glass RPCs at the Puy de D\^ome volcano~\cite{Fehr2012}, 50$\times$50~cm$^2$ Micromegas detectors at the Khufu Pyramid~\cite{Morishima2017}, multi-wire proportional chambers at the Sakurajima volcano~\cite{Olah2018}, and 1~m$^2$ double-gap glass RPCs in the Lousal mine~\cite{Sarmento2024}.

In this paper, we present the development and characterization of three portable, gas-tight glass RPC prototypes, fabricated by our collaboration as tracking layers for a compact and lightweight muon telescope. This work extends previous developments of gas-tight glass RPCs within our collaboration~\cite{Basnet2020,Samalan2024,Kumar2025,Abhishek2025,Ikram2025,Karnam2026}. The detectors are designed to operate in sealed mode or at very low gas flow and to be read out with a low-power DAQ, so that the system can be deployed autonomously in the field without the need for an on-site gas mixing unit. Particular attention is given to their mechanical robustness, operational safety, and low-cost modular construction.\

The paper is organized as follows: Section~\ref{sec:prototypes} describes the design and construction of the three prototypes, including the electrode coating, readout boards, gas enclosures and data-acquisition systems. Section~\ref{sec:performance} reports their detection efficiency, time and spatial resolution, and long-term performance stability, measured with cosmic-ray muons in the laboratory. Section~\ref{sec:absorption} presents a proof-of-principle absorption measurement with lead blocks as a first illustration of the imaging capability of the system. Finally, Section~\ref{sec:summary} summarizes the results and outlines future developments.

\section{Portable gRPC Prototypes}\label{sec:prototypes}
\subsection{Overview of the three prototypes}\label{sec:overview}
Three small-area glass RPC prototypes, labelled A, B and C, have been developed at UCLouvain, NISER and VUB, respectively~\cite{Kumar2025,Abhishek2025,Samalan2024,Ikram2025}. They share the same design goals: operation with low or zero gas flow (i.e.\ sealed systems), a lightweight, autonomous and modular construction suitable for confined or remote locations. The main characteristics of the three prototypes are summarized in Table~\ref{tab:prototypes}. All chambers use float-glass electrodes and are operated in avalanche mode with the standard gas mixture C$_2$H$_2$F$_4$/i-C$_4$H$_{10}$/SF$_6$ (95.2/4.5/0.3). Prototype A is fully operated in sealed mode, whereas prototypes B and C are designed for sealed operation but are currently operated in an open-loop gas configuration.

\begin{table*}[t]
\tbl{Main characteristics of the three gRPC prototypes.\label{tab:prototypes}}{%
\begin{tabular}{llll}
\toprule
 & Prototype A & Prototype B & Prototype C \\
\colrule
Electrode material / thickness & Float glass / 1.1~mm & Float glass / 3~mm & Float glass / 1.1~mm \\
Resistive coating & Serigraphy ($\sim$0.2--0.6~M$\Omega$/$\square$), & Spray gun ($\sim$1~M$\Omega$/$\square$) & Spray gun ($\sim$1.5~M$\Omega$/$\square$) \\
 & hand-painted (0.5--1~M$\Omega$/$\square$) & & \\
Electrode active area & 16$\times$16~cm$^2$ & 16$\times$16~cm$^2$ & 28$\times$28~cm$^2$ \\
Gas gap & 1~mm, single gap & 2~mm, single gap & 1~mm, double gap \\
Readout strips /strip pitch & 16 - 1D / 1~cm & 16$\times$16 - 2D / 1~cm & 32$\times$32 - 2D / 0.8~cm \\
Outer casing & Aluminium box & Acrylic box & Gas volume closed by top \\
 & & & and bottom PCBs \\
Gas mixture & \multicolumn{3}{c}{C$_2$H$_2$F$_4$/i-C$_4$H$_{10}$/SF$_6$ = 95.2/4.5/0.3} \\
DAQ & \multicolumn{3}{c}{NIM and ASIC + FPGA} \\
Sealed mode & Yes & Yes (currently open loop) & Yes (currently open loop) \\
\botrule
\end{tabular}}
\end{table*}

\subsection{Detector layouts}\label{sec:layouts}
Several small-area glass RPC prototypes were fabricated over the years. Three of them are shown in Fig.~\ref{fig:layouts}. Their design evolved in three main directions: from a single-gap to a double-gap configuration to increase the detection efficiency, from one-dimensional (1D) to two-dimensional (2D) strip readout boards to enable tracking reconstruction, and from a large external gas box to an internal, cable-free gas volume, which makes the detector more lightweight and improves the gas tightness.

In prototype~A (Fig.~\ref{fig:layouts}, left), the detector is a single-gap chamber with an active area of 16$\times$16~cm$^2$, built from 1.1~mm thick glass electrodes. The 1~mm gas gap is set by circular spacers. The signal is read out by a PCB with 16 strips at 1~cm pitch, providing a one-dimensional (1D) position measurement. The glass stack and the readout board are placed inside a sealed aluminium gas box, and the detector is operated in sealed mode~\cite{Kumar2025}.

\begin{figure}[H]
\centering
\includegraphics[width=0.65\columnwidth]{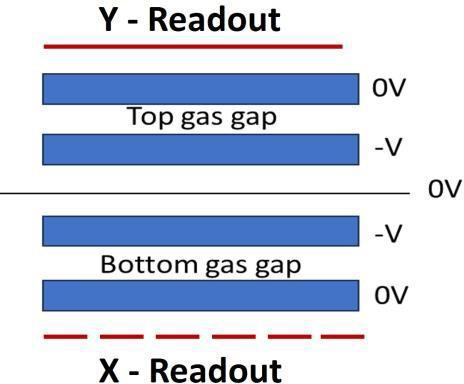}
\caption{Schematic layout of prototype B module.}
\label{fig:schemeB}
\end{figure}

Prototype~B (Fig.~\ref{fig:layouts}, middle) follows the same approach, with the
glass stack and readout PCBs housed in an airtight acrylic casing. Each module has an
active area of 16$\times$16~cm$^2$ and consists of two single-gap RPCs, made
of 3~mm thick glass electrodes with a 2~mm gas gap. The glass electrodes are placed between two orthogonal readout planes with 16 strips at 1~cm pitch, which give the $X$ and $Y$ coordinates of the muon hit (Fig.~\ref{fig:schemeB}). The detector is designed for sealed operation but is currently operated with gas flow~\cite{Abhishek2025}.

\begin{figure}[H]
\centering
\includegraphics[width=\columnwidth]{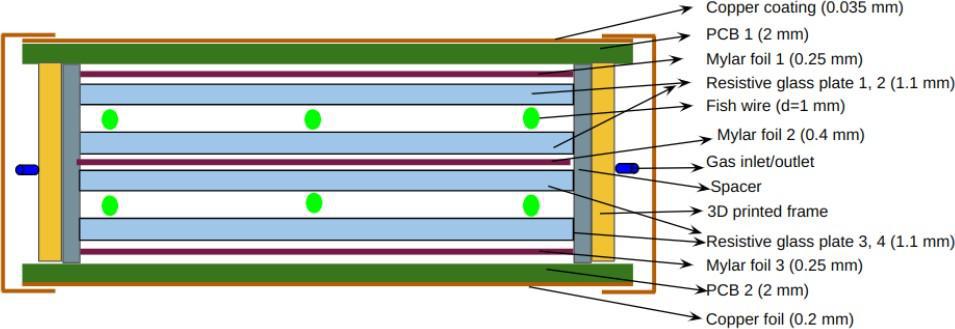}
\caption{Schematic cross-section of prototype C. The glass electrode stack is insulated by mylar foils and enclosed by a 3D-printed frame between the top and bottom readout PCBs.}
\label{fig:stackC}
\end{figure}

In prototype~C (Fig.~\ref{fig:layouts}, right), the external gas box is removed.
The detector is a double-gap chamber with an active area of 28$\times$28~cm$^2$,
built from four 1.1~mm-thick glass electrodes. Each 1~mm gas gap is set by nylon fishing wires of 1~mm diameter, fixed to 3D-printed spacer blocks. The two gaps are stacked with their HV electrodes facing each other, separated by an insulating Mylar foil. The gas volume is defined by a single-piece 3D-printed frame and closed on the top and bottom sides by the readout PCBs, sealed with O-rings (Fig.~\ref{fig:stackC}). The frame has openings only for the gas inlet and outlet. The HV, ground, and signal connections, as well as the readout of a pressure, temperature, and humidity (PTH) sensor placed inside the gas volume, are routed through traces embedded in the PCBs, so that no cables pass through the frame. The PTH data are used to correct the applied voltage for pressure and temperature variations.

\subsection{Detector construction}\label{sec:construction}
\subsubsection{Glass electrode coating}
The glass electrodes are coated with a semi-conductive paint obtained by mixing the conductive paint Electrodag~6017SS with the resistive paint PM-404. Several coating methods have been used: manual paint roller, manual spray painting, and manual and automated serigraphy (silk-screen printing). The resulting surface resistivities range between 0.2 to 1.5~M$\Omega$/$\square$. The long-term stability of the coating was monitored for both serigraphy and spray coating techniques.

\begin{figure}[H]
\centering
\includegraphics[width=0.65\columnwidth]{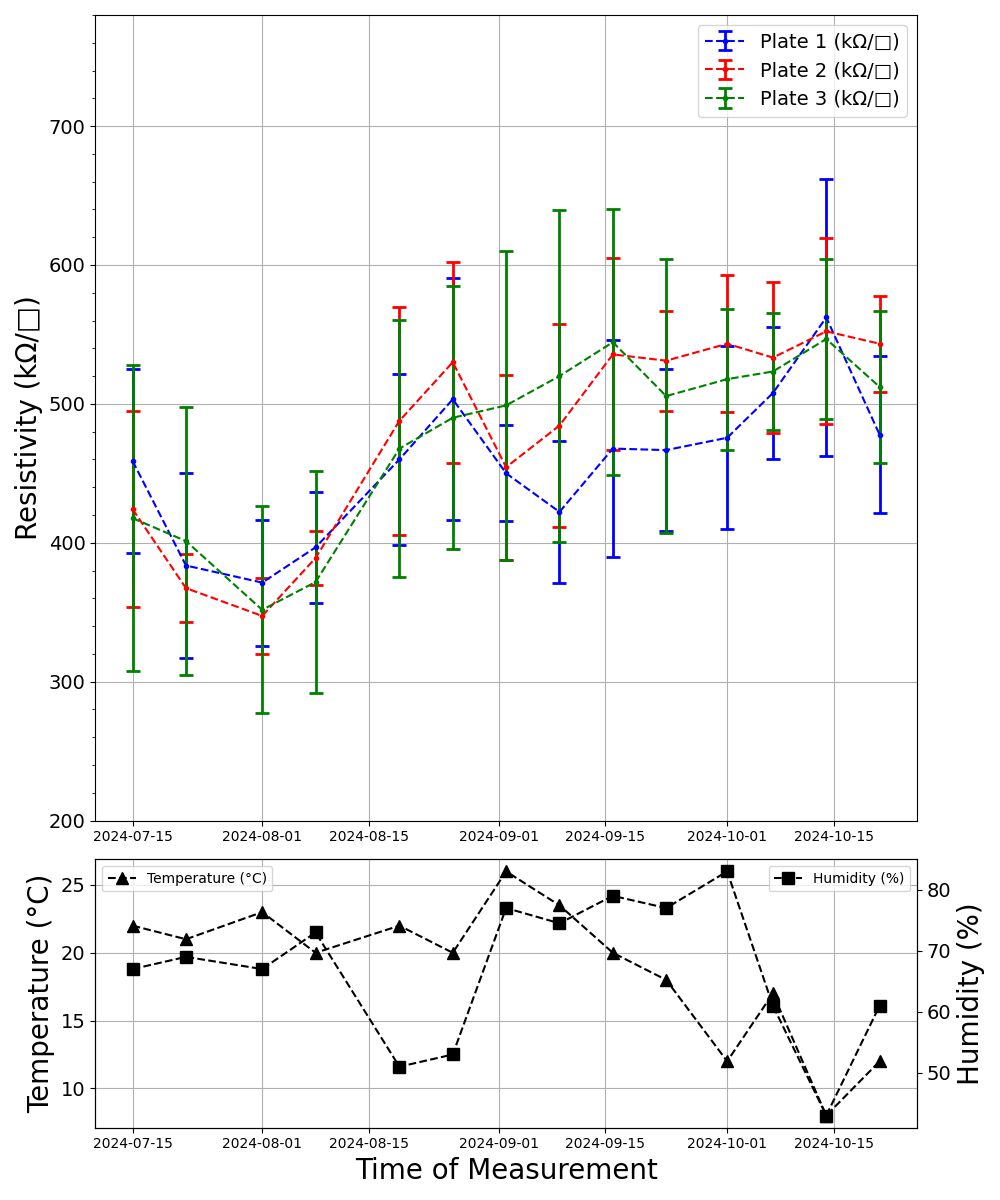}
\caption{Surface resistivity (k$\Omega$/$\square$) of three serigraphy-coated glass plates of prototype A as a function of time (top), with the ambient temperature and relative humidity (bottom).}
\label{fig:coatingA}
\end{figure}

The electrodes of prototype~A were coated by serigraphy, and the surface resistivity of three plates was monitored from July to October 2024, together with the ambient temperature and relative humidity (Fig.~\ref{fig:coatingA}). The three plates follow the same trend and agree within the measurement uncertainties. Their resistivity decreases from about 420--460~k$\Omega$/$\square$ to a minimum of about 350--375~k$\Omega$/$\square$ in early August, and then rises gradually to about 500--560~k$\Omega$/$\square$ by mid-October. During this period, the temperature varied between about 12 and 26~$^\circ$C and the relative humidity between about 45 and 80\%.

\begin{figure}[H]
\centering
\includegraphics[width=0.7\columnwidth]{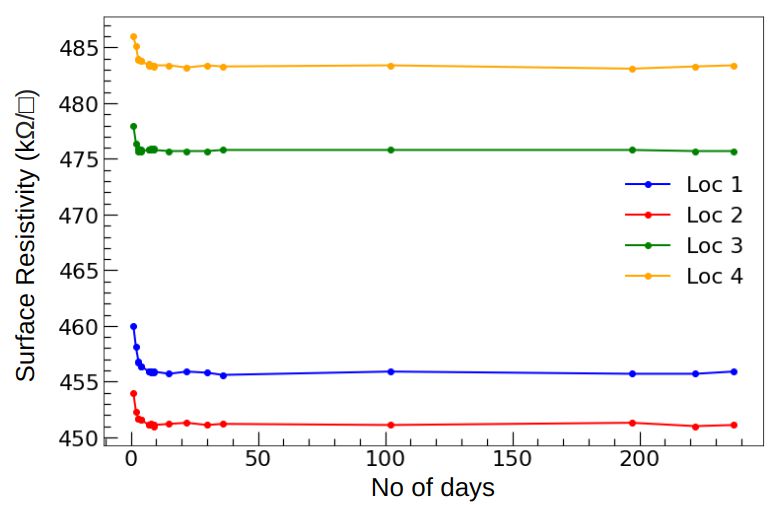}
\caption{Surface resistivity of a spray-coated glass plate of prototype C, measured at four locations as a function of time.}
\label{fig:coatingC}
\end{figure}

A similar long-term monitoring was carried out for the electrodes of prototype~C, which were coated by manual spray painting. The surface resistivity was then measured at four locations on one of the glass plates over about 240 days (Fig.~\ref{fig:coatingC}). During the first few days, the values decrease slightly and then remain stable for the rest of the period. The values at the four locations lie between about 451 and 484~k$\Omega$/$\square$, which indicates a uniform coating across the electrode surface.

\begin{figure*}[t]
\centering
\begin{minipage}[b]{0.30\textwidth}
\centering
\includegraphics[width=\linewidth,height=4.2cm,keepaspectratio]{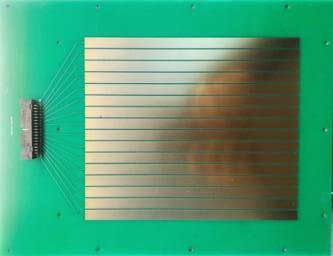}
\end{minipage}\hfill
\begin{minipage}[b]{0.30\textwidth}
\centering
\includegraphics[width=\linewidth,height=4.2cm,keepaspectratio]{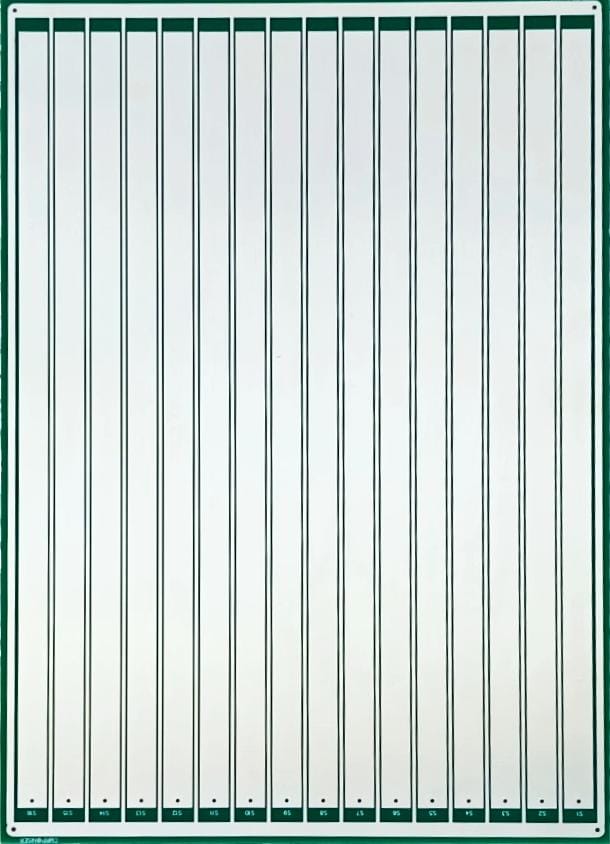}
\end{minipage}\hfill
\begin{minipage}[b]{0.34\textwidth}
\centering
\includegraphics[width=\linewidth,height=4.2cm,keepaspectratio]{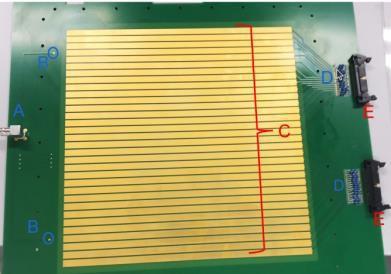}
\end{minipage}
\caption{Readout boards of the three prototypes. Left: prototype~A board with 16 copper strips at 1~cm pitch and integrated signal traces. Middle: prototype~B board with 16 copper strips at 1~cm pitch and signal cables soldered to the strips. Right: prototype~C PCB with 32 readout strips of 0.8~cm width (C), BNC connector for the HV supply (A), HV and ground traces to the electrodes (B), 50~$\Omega$ termination resistors (D) and flat-cable connectors for signal collection (E).}
\label{fig:readout}
\end{figure*}

\subsubsection{Readout boards}
The readout boards of prototype A consist of a PCB with 16 embedded copper strips of 1~cm width, from which copper traces route the induced signals to the front-end electronics (Fig.~\ref{fig:readout}, left). To improve the spatial granularity, these boards are currently being upgraded to a new design with 64 readout strips.

Prototype B uses PCB-based boards with 16 copper strips at a pitch of 1~cm, with cables soldered directly to the strips to transfer the signals to the readout electronics (Fig.~\ref{fig:readout}, middle). 

Each PCB of prototype~C (Fig.~\ref{fig:readout}, right) integrates 32 readout strips of 0.8~cm pitch, and the two PCBs are oriented orthogonally to provide 2D readout. Signal traces route the strips to two 17-pin flat-cable connectors, and a solder mask over the traces reduces cross-talk. Dedicated slots for 50~$\Omega$ termination resistors limit signal reflections, and a BNC connector supplies the HV. The back of each PCB is coated with a 35~$\mu$m copper layer. After assembly, copper foil connects the top and bottom PCBs to form a Faraday cage.

\subsubsection{Gas tightness and leak test}

A key requirement for portable glass RPCs is to keep the gas volume small and leak-tight, so that the detectors can be operated in sealed mode. The three prototypes follow different approaches to meet this requirement.

\begin{figure}[H]
\centering
\includegraphics[width=0.7\columnwidth]{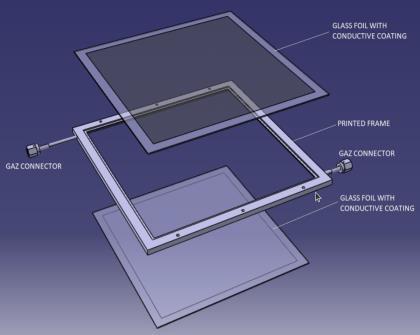}
\caption{Exploded view of the 3D-printed frame under development for prototype A.}
\label{fig:frameA}
\end{figure}

Prototype A is currently operated with a sealed aluminium gas enclosure, which provides the required gas tightness. For future developments, the 3D-printed frame solution is being adapted to this prototype in order to reduce the gas volume (Fig.~\ref{fig:frameA}). In this design, which is still under development, the printed frame holds the two coated glass electrodes at a fixed gap and integrates the gas inlet and outlet connectors.

\begin{figure}[H]
\centering
\includegraphics[width=0.5\columnwidth]{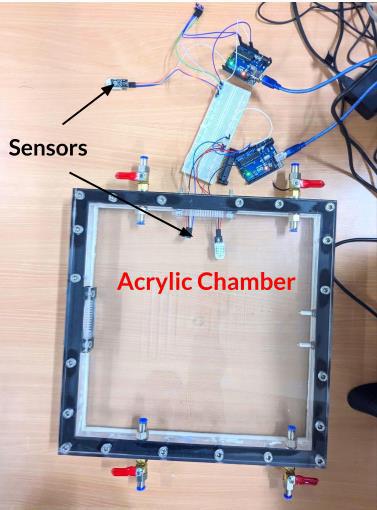}
\caption{Acrylic (PMMA) chamber of prototype B, equipped with DHT22 sensors for monitoring the relative humidity and temperature inside the casing.}
\label{fig:acrylicB}
\end{figure}

In prototype B, the glass stack and the readout panels are housed in an acrylic (PMMA) chamber (Fig.~\ref{fig:acrylicB}). The relative humidity and temperature inside the chamber are monitored with DHT22 sensors read out by an Arduino UNO R3 board~\cite{Abhishek2025}. One sensor is placed on the inner floor of the casing and a second one outside as reference.

\begin{figure}[H]
\centering
\includegraphics[width=0.7\columnwidth]{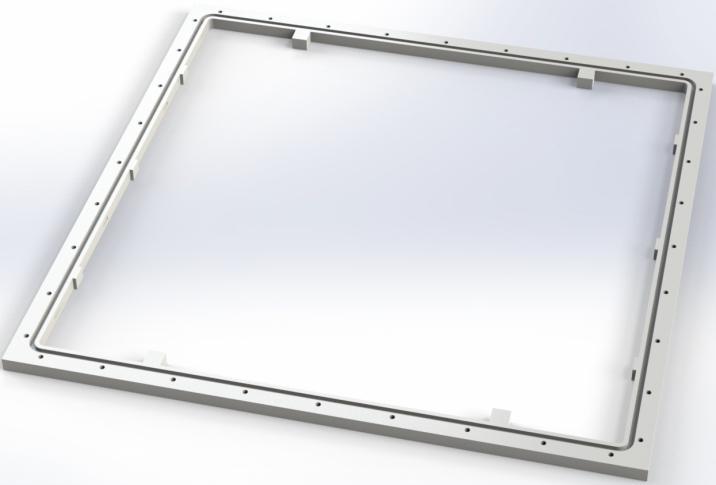}
\caption{Design of the 3D-printed frame of prototype C.}
\label{fig:frameC}
\end{figure}

In prototype C, the gas volume is defined by a single-piece 3D-printed frame that encloses the glass stack and is screwed to the top and bottom PCBs, which close the gas volume (Fig.~\ref{fig:frameC}). Gas distribution slots built into the frame ensure a uniform gas flow, O-rings provide the gas tightness, and 3D-printed spacer blocks fix the fishing wires and hold the glass stack in place. All electrical connections are made through traces embedded in the PCBs, so that no cables enter the gas volume. The design is also glue-free, which allows the detector to be reopened in case of failure or for reconfiguration. The temperature, pressure and humidity of the gas are monitored with internal sensors. 

\begin{figure}[H]
\centering
\includegraphics[width=\columnwidth]{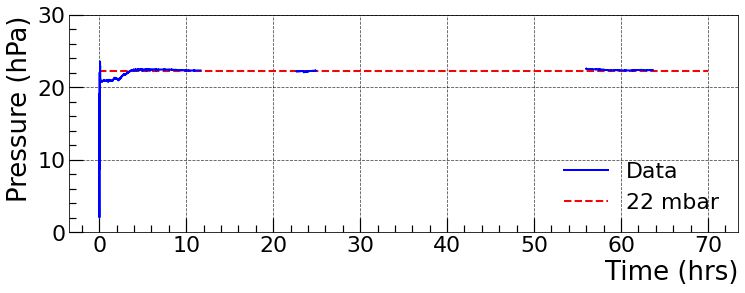}
\caption{Gas leak test of prototype C: internal pressure as a function of time for the chamber filled with argon at an overpressure of 22~mbar (dashed line).}
\label{fig:leakC}
\end{figure}

The internal pressure sensor was also used to verify the gas tightness of this design. For this leak test, the chamber was filled with argon at an overpressure of 22~mbar and the pressure was recorded over about 65 hours (Fig.~\ref{fig:leakC}). After the initial filling, the pressure stabilizes at the set value and no measurable decrease is observed over the whole monitoring period, confirming the leak tightness of the 3D-printed frame.

\subsection{Data acquisition and power supplies}\label{sec:daq}

Prototype A is read out with front-end boards developed for the CMS muon RPC system, which combine charge-sensitive preamplifiers and discriminators~\cite{Abbrescia2000}. Four front-end ASICs, serving four readout boards with 16 strips each, are connected via I$^2$C to an FPGA and CPU module (Trenz Electronic TE0720-03-1CF), which handles the data acquisition, controls the high-voltage (HV) supply, and stores the data on an SD card; the system can be accessed remotely via Ethernet or WiFi. The HV could be provided by a compact ISEG DPS DC/DC module with a single output channel, delivering up to 10~kV at a maximum current of 1~mA from a 24~V input.

Prototype B is read out with the CAEN DT5550W DAQ platform, which combines a programmable FPGA with a PETIROC-2A ASIC piggyback board that is currently being tested~\cite{CAENDT5550W}. Each PETIROC-2A ASIC provides 32 input channels, so that four ASICs read out 128 channels in total, with positive or negative signal polarity, a dynamic range from 160~fC to 480~pC, 32 trigger outputs, and a TDC with 40~ps bin width. All RPCs are powered by a portable CAEN A7512DN single-channel HV module, which provides up to 12~kV with a maximum current of 20~$\mu$A, and voltage and current monitoring resolutions of 200~mV and 500~pA, respectively.

Prototype~C was tested in the laboratory with a CAEN V792 charge-to-digital converter (QDC). The strip signals are sent through flat cables and CAEN A992 impedance adapters, for 50~$\Omega$ matching, to the QDC, which records the signals of all channels for every trigger. The chamber was powered with CAEN HV modules~\cite{Samalan2024}.
\section{Performance Tests in the Laboratory}\label{sec:performance}
The performance of the three prototypes was characterized in the laboratory with cosmic-ray muons, using plastic scintillators coupled to photomultiplier tubes (PMTs) as an external trigger~\cite{BASIRI2026104692}.

\subsection{Prototype A}\label{sec:perfA}
\begin{figure}[H]
\centering
\includegraphics[width=0.65\columnwidth]{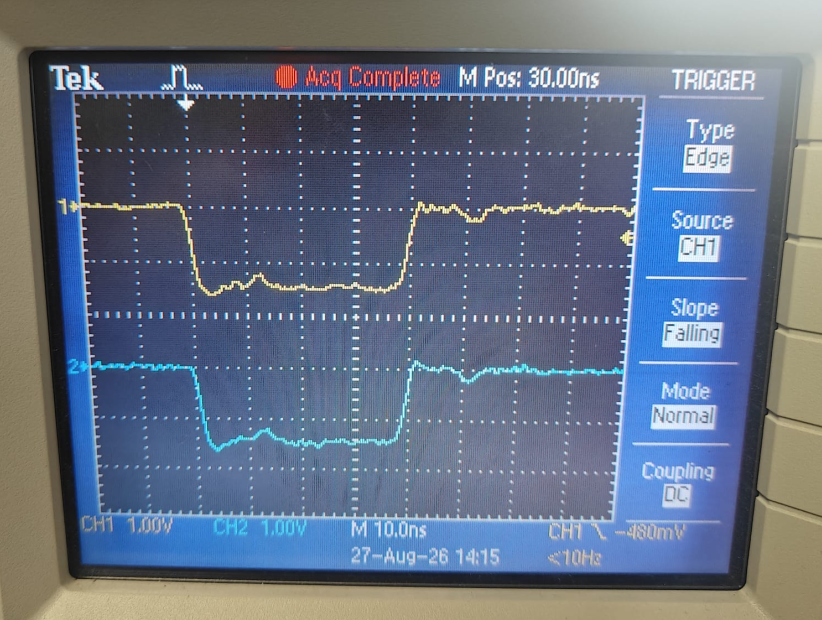}
\caption{Scintillator NIM pulses forming the coincidence trigger (10 ns/div).}
\label{fig:Coincidence_window}
\end{figure}

\begin{figure}[H]
\centering
\includegraphics[width=\columnwidth]{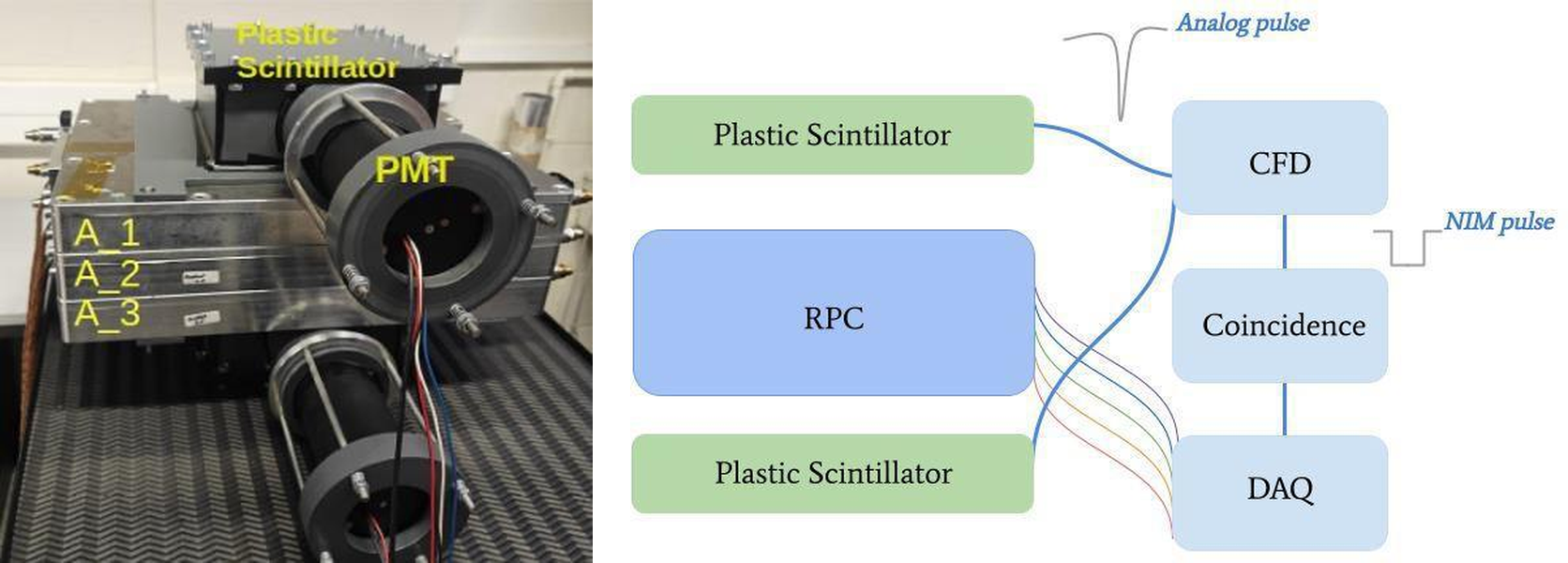}
\caption{Cosmic-ray test setup of prototype A. Left: chambers A$_1$--A$_3$ stacked between two plastic scintillators. Right: readout scheme; the scintillator signals pass through a CFD, and their coincidence triggers the DAQ to record the RPC signals.}
\label{fig:setupA}
\end{figure}

Three identical chambers of prototype A (A$_1$--A$_3$) were stacked between two plastic scintillators with the same active area as the RPCs (Fig.~\ref{fig:setupA}). The signals from the two scintillators, each coupled to a photomultiplier tube (PMT), are shaped by a quad timer NIM module and combined using a logical AND to form the trigger. A coincidence requires temporal overlap between the two logic pulses, whose widths determine the coincidence acceptance window. The oscilloscope traces show negative pulses approximately 40 ns wide at a horizontal scale of 10 ns per division (Fig.~\ref{fig:Coincidence_window}). The leading edge of the coincidence output defines the trigger time and corresponds to a falling voltage edge in negative NIM logic. The detection efficiency is defined in Eq.~\eqref{eq:eff}. 

\begin{equation}
\varepsilon=\frac{N(\mathrm{RPC~hit~after~filters}\wedge\mathrm{scintillator~coincidence})}{N(\mathrm{scintillator~coincidence})}\times 100\,
\label{eq:eff}
\end{equation}
where the filters reject noise hits outside the RPC time window and events with large strip multiplicity~\cite{Ikram2025}.

\begin{figure}[H]
\centering
\includegraphics[width=0.85\columnwidth]{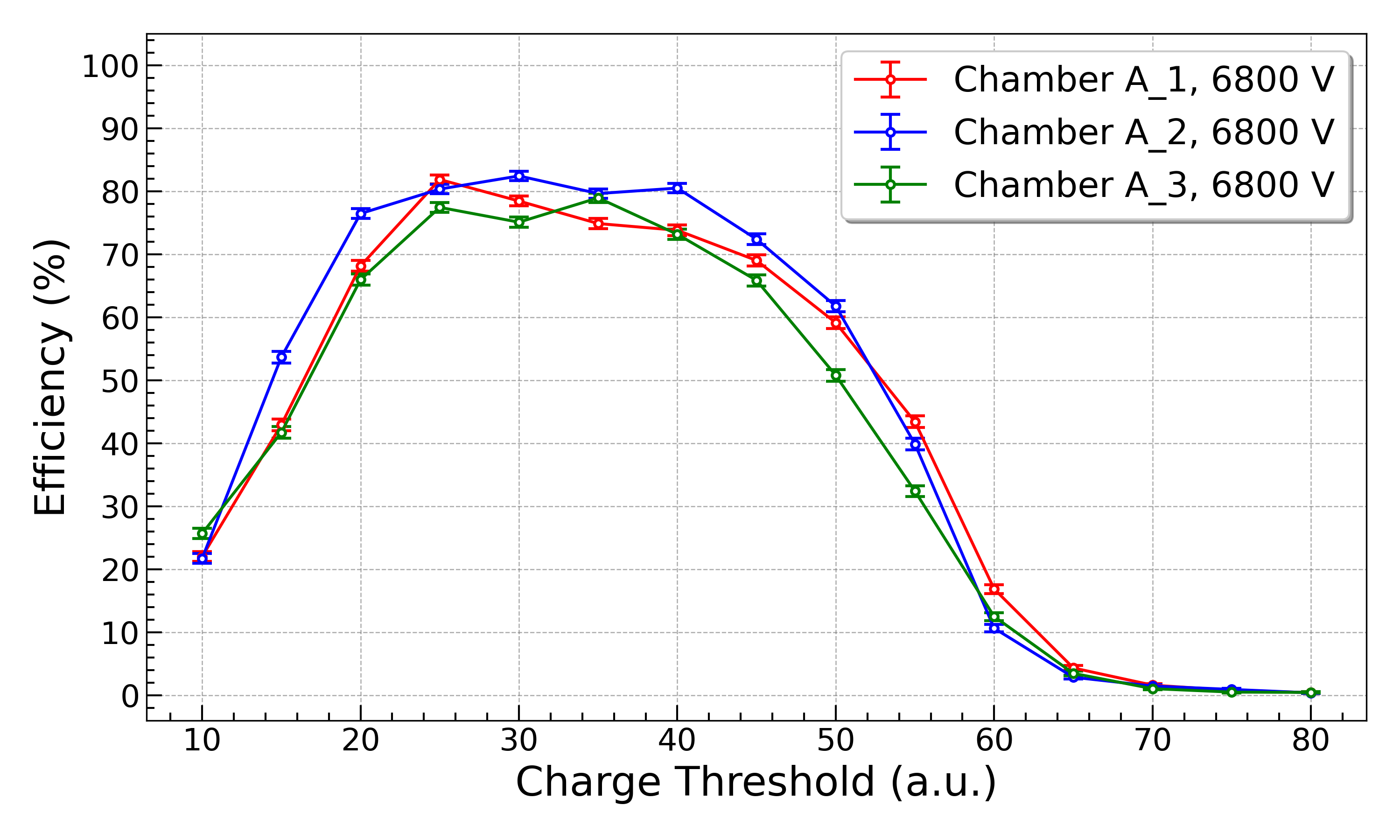}
\caption{Detection efficiency of the three prototype A chambers as a function of the charge threshold, for chambers A$_1$, A$_2$ and A$_3$ at 6.8~kV, after two years of operation.}
\label{fig:effThrA}
\end{figure}

\begin{figure}[H]
\centering
\includegraphics[width=0.82\columnwidth]{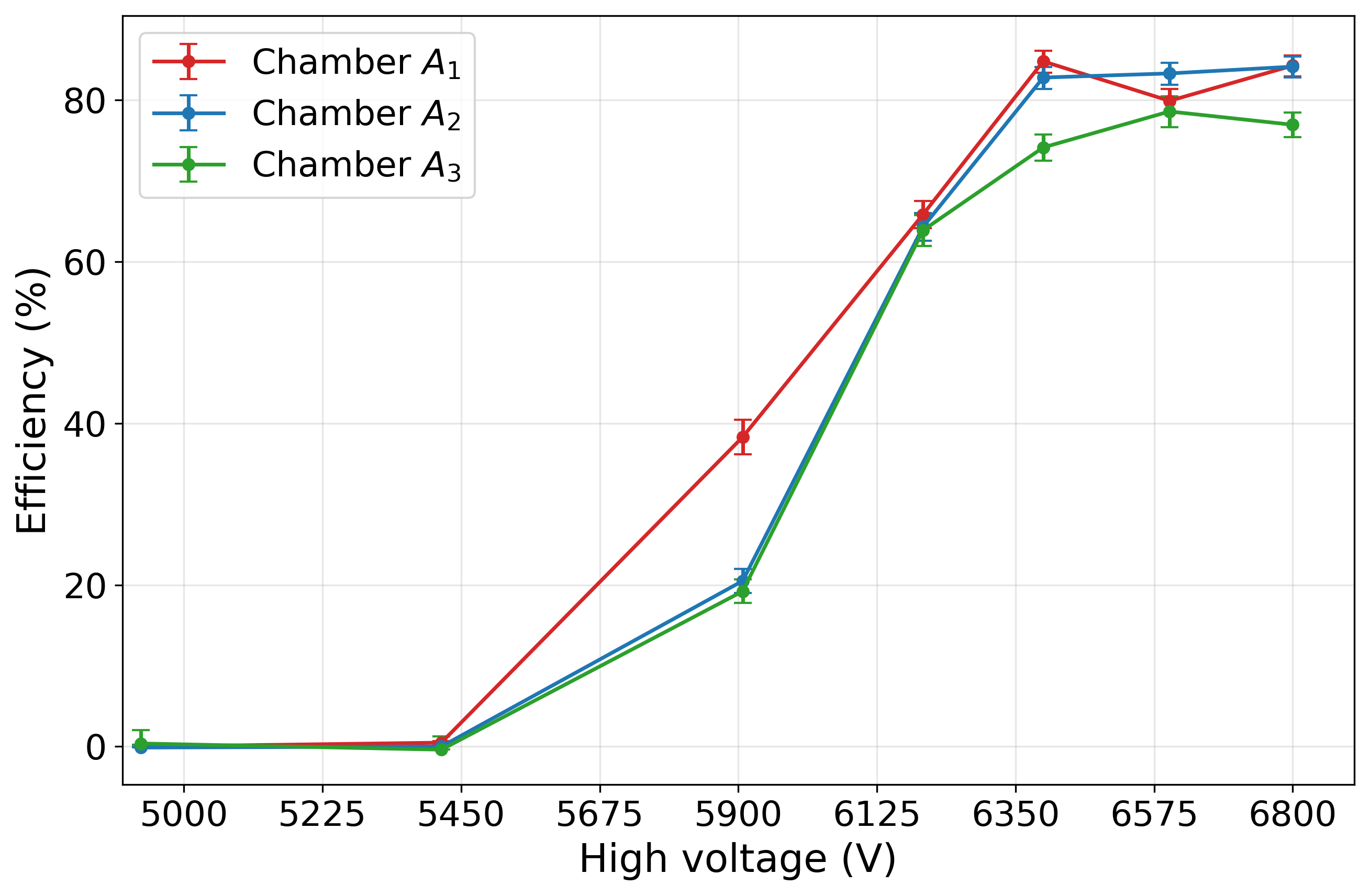}
\caption{Detection efficiency of the three prototype A chambers as a function of the high voltage.}
\label{fig:AeffHV}
\end{figure}

The efficiency of the three chambers, computed using Eq.~\eqref{eq:eff}, was measured as a function of the charge threshold at HV values of 6.8~kV (Fig.~\ref{fig:effThrA}), and as a function of high voltage (Fig.~\ref{fig:AeffHV}) after two years of operation, with statistical uncertainties estimated using binomial statistics. All three RPC chambers show a broad efficiency maximum around 25–40 (a.u.), with peak efficiencies of approximately 78–83\%. and the efficiency decreases with increasing threshold above this value. Chamber A$_2$ has the most stable response in this range. At higher thresholds, the efficiency decreases sharply. The three chambers show a consistent behaviour, with a higher efficiency at 6.8~kV, which was therefore chosen as the working point. 

\begin{figure}[H]
\centering
\includegraphics[width=0.9\columnwidth]{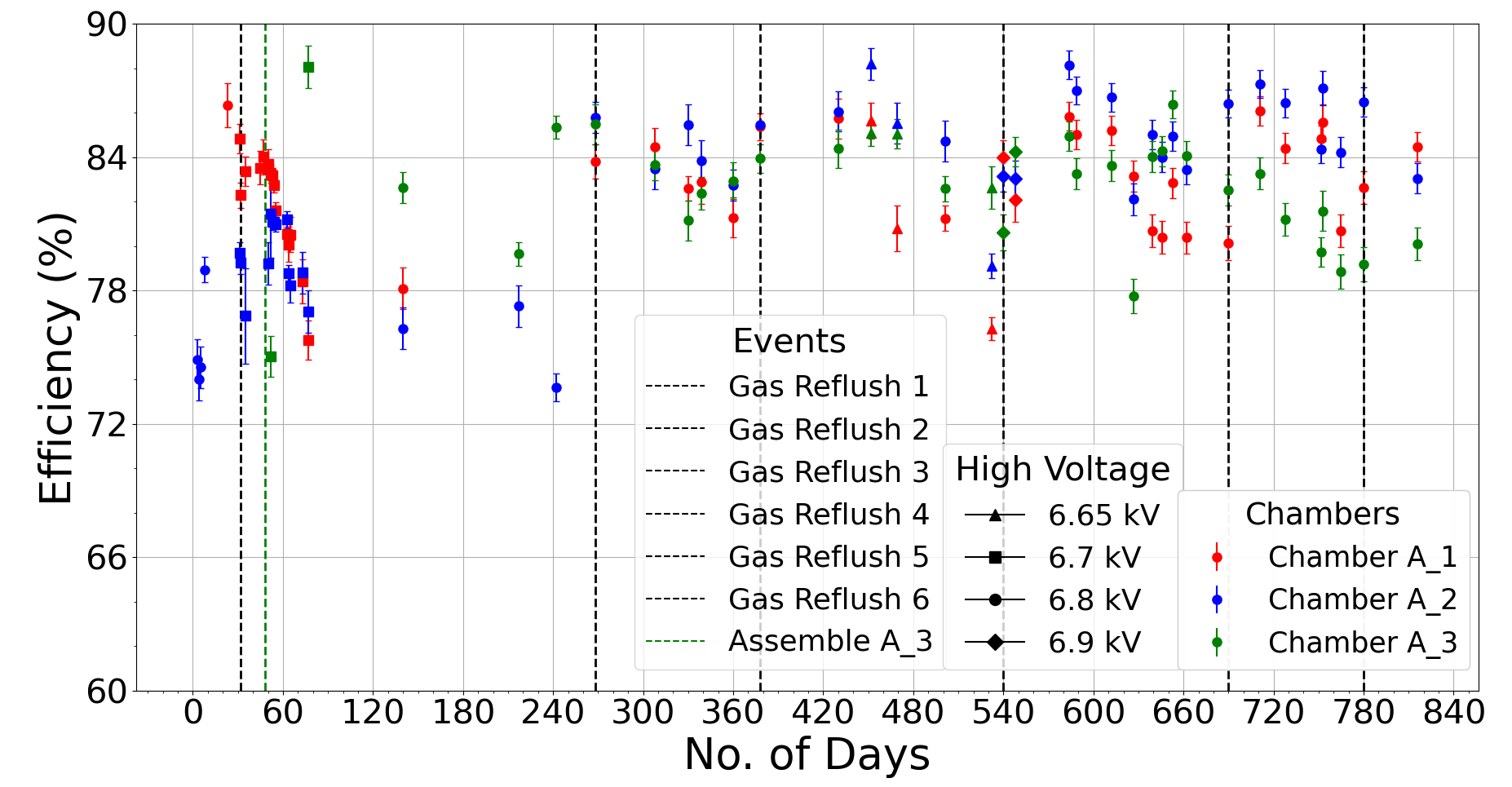}
\caption{Detection efficiency of the three prototype A chambers as a function of time over more than two years. Vertical lines indicate the gas refills of chamber A$_3$; marker shapes indicate the applied high voltage.}
\label{fig:effTimeA}
\end{figure}

The long-term stability of the chambers operated in sealed mode was monitored over more than two years at HV values between 6.65 and 6.9~kV (Fig.~\ref{fig:effTimeA}). A gradual decrease of the efficiency is observed between gas refills, and periodic flushing of the chambers restores their performance, so that the efficiency remains stable at about 80--85\% over the whole period.

\subsection{Prototype B}\label{sec:perfB}
\begin{figure}[H]
\centering
\includegraphics[width=0.85\columnwidth]{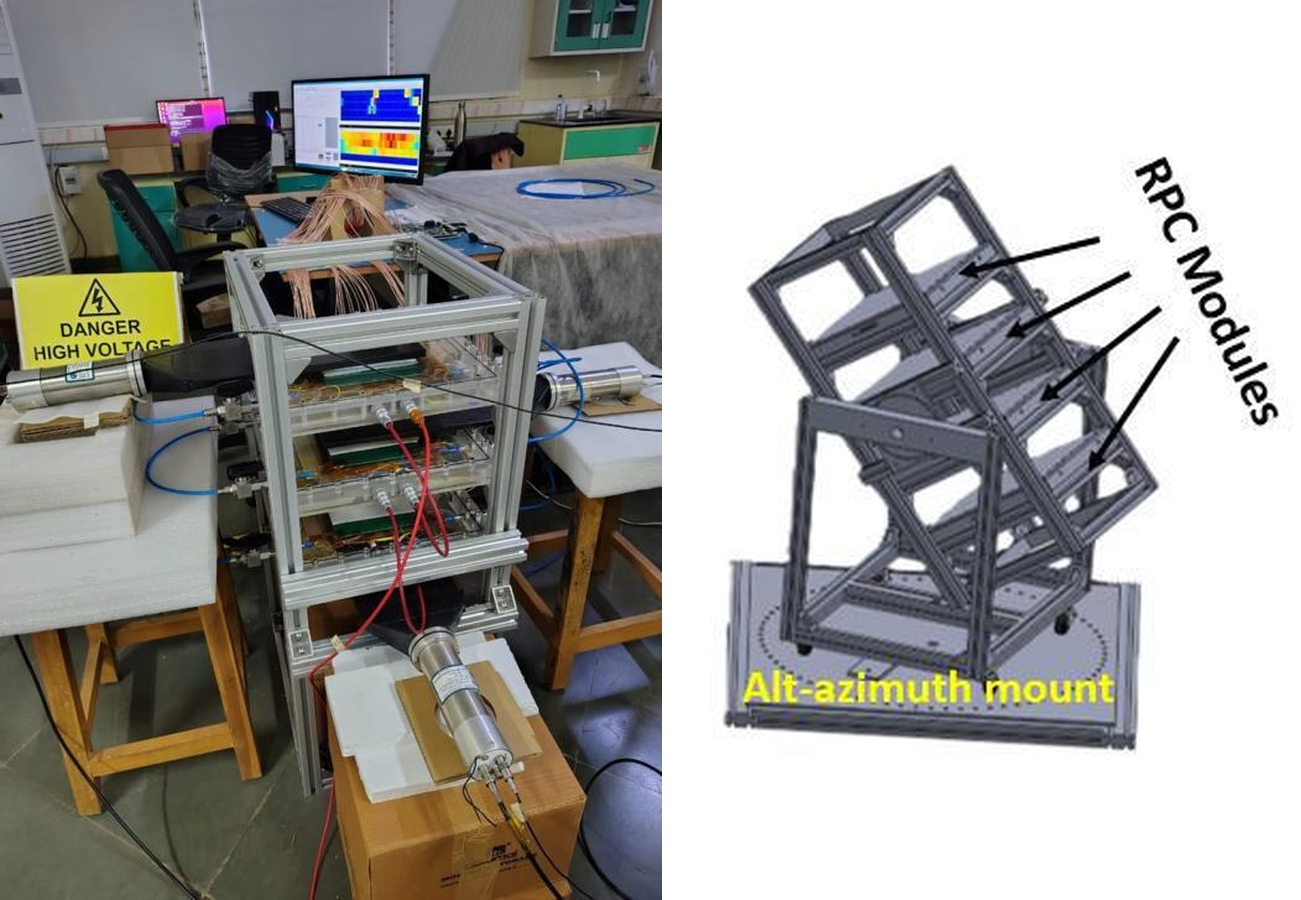}
\caption{Three-layer muon telescope based on prototype B. Left: telescope in the laboratory with the trigger scintillators and PMTs. Right: design of the alt-azimuthal mount holding the RPC modules.}
\label{fig:setupB}
\end{figure}

A three-layer muon telescope was built with standalone glass RPC modules of prototype B (Fig.~\ref{fig:setupB}). Each module houses two RPCs sandwiched between two orthogonal strip readout boards, enabling two-dimensional ($X$, $Y$) tracking. The modules are mounted on an alt-azimuth platform, which allows the telescope to be oriented towards the target~\cite{Abhishek2025}.

\begin{figure}[H]
\centering
\includegraphics[width=0.72\columnwidth]{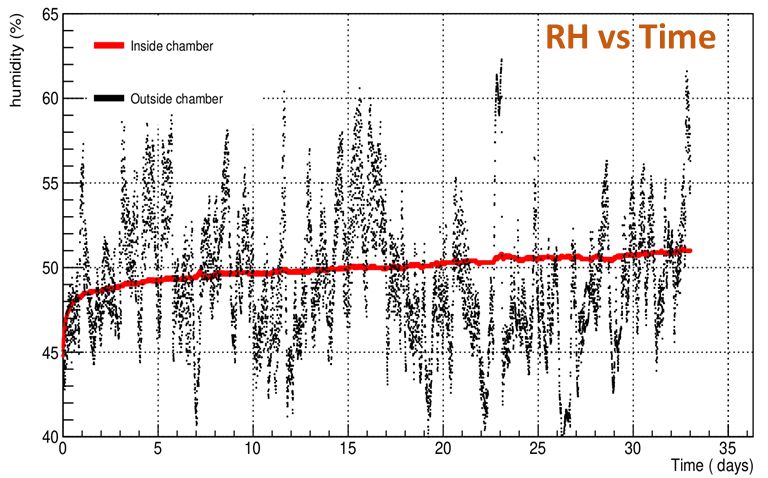}
\caption{Relative humidity inside (red) and outside (black) the acrylic chamber of prototype B as a function of time.}
\label{fig:RHB}
\end{figure}

The relative humidity (RH) inside the acrylic chamber was monitored over more than one month (Fig.~\ref{fig:RHB}). While the ambient humidity fluctuates between about 40\% and 65\%, the RH inside the chamber remains stable at about 48--51\%(a slow residual rise of a few percent), showing that the acrylic casing strongly damps ambient fluctuations, with a very small moisture ingress.

\begin{figure}[H]
\centering
\includegraphics[width=0.9\columnwidth]{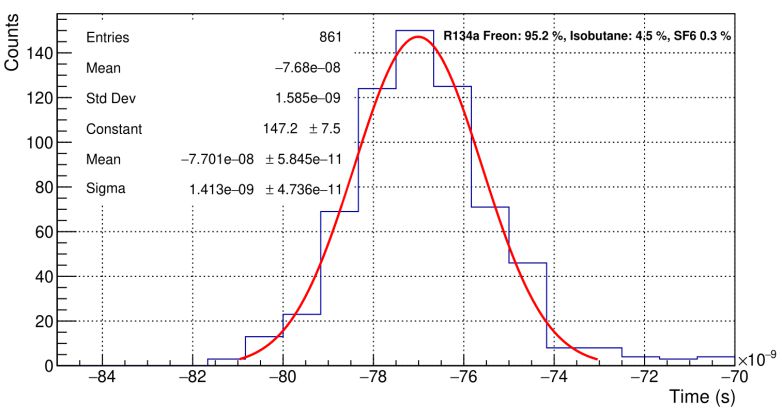}
\caption{Distribution of the time difference between the trigger and the RPC signal for prototype B, with a Gaussian fit (red) giving a time resolution of about 1.4~ns.}
\label{fig:timeB}
\end{figure}

The time resolution of the RPCs was obtained from the distribution of the RPC signal time measured relative to the scintillator trigger (Fig.~\ref{fig:timeB}). A Gaussian fit gives a standard deviation of about 1.4~ns. This value includes the time jitter of the scintillator trigger and the readout electronics, so the intrinsic time resolution of the RPC is better.

\begin{figure}[H]
\centering
\includegraphics[width=0.9\columnwidth]{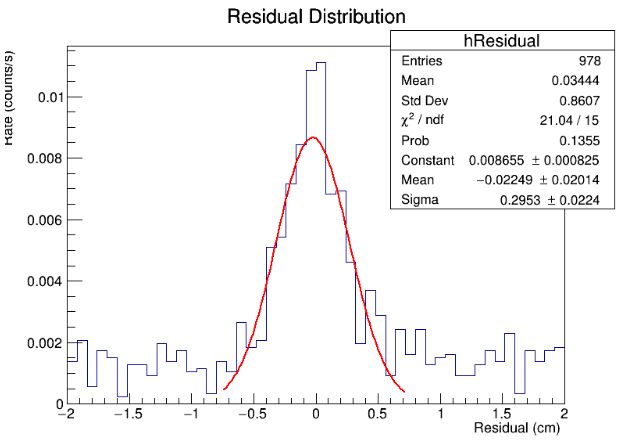}
\caption{Distribution of the track residuals for prototype B, with a Gaussian fit (red) giving a position resolution of about 3~mm.}
\label{fig:posB}
\end{figure}

The strip multiplicity is low: for most cosmic-ray muon events, the avalanche charge is confined to one or two strips~\cite{Karnam2026}. The position resolution was determined from the distribution of the track residuals (Fig.~\ref{fig:posB}), and a Gaussian fit yields a standard deviation of about 3~mm. This matches the binary limit of the 1~cm strip pitch ($10~\mathrm{mm}/\sqrt{12} \approx 2.9$~mm).

\begin{figure}[H]
\centering
\includegraphics[width=0.8\columnwidth]{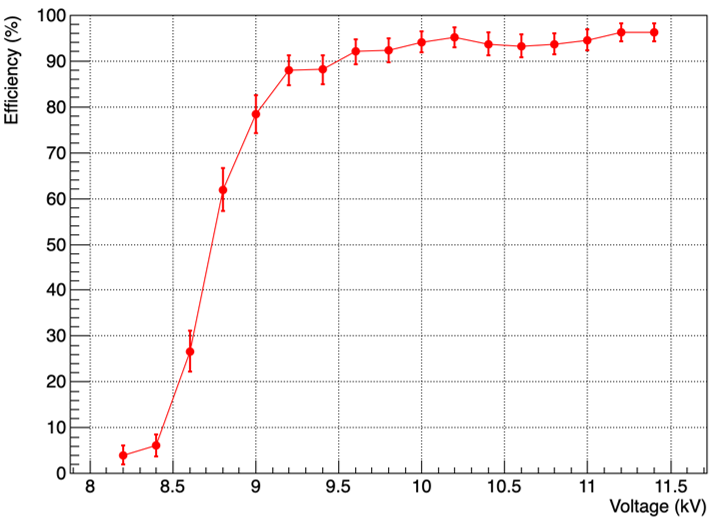}
\caption{Cosmic-ray muon detection efficiency of prototype B as a function of the applied voltage.}
\label{fig:effB}
\end{figure}

The detection efficiency as a function of the applied voltage is shown in Fig.~\ref{fig:effB}. The efficiency rises steeply above 8.5~kV and reaches a plateau of at least 95\% for applied voltages above 10~kV.

\subsection{Prototype C}\label{sec:perfC}
\begin{figure}[H]
\centering
\includegraphics[width=1.15\columnwidth]{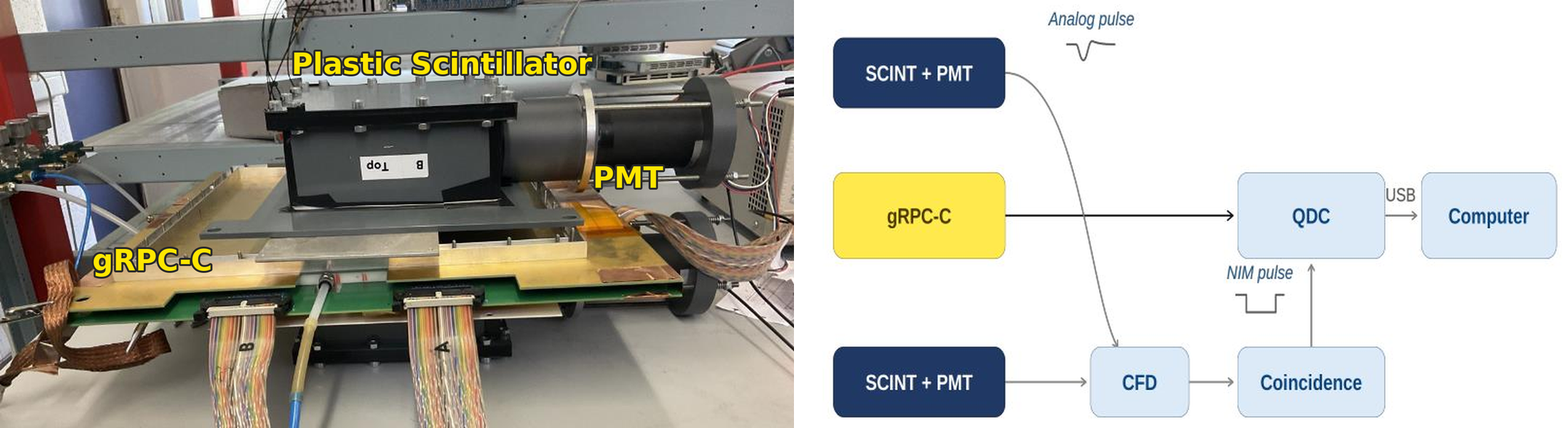}
\caption{Cosmic-ray test setup of prototype C. Top: double-gap chamber placed between two trigger scintillators read out by PMTs. Bottom: schematic of the NIM/VME-based trigger and data-acquisition chain.}
\label{fig:setupC}
\end{figure}

The double-gap prototype C was positioned between two 16$\times$16~cm$^2$ trigger scintillators, so that only 18 of the 32 strips of the readout plane were covered during the measurements (Fig.~\ref{fig:setupC}). The PMT signals are discriminated by a CFD and put in coincidence to trigger a charge-to-digital converter (QDC) that records the RPC signals, which are then transferred to a computer via USB. For these first performance studies, only one readout PCB was used, with its 32 strips connected to the 32 channels of the QDC. Before each cosmic run, a pedestal run was taken, and the threshold of each channel was set from its pedestal distribution; only strips with a signal above threshold are used in the analysis. The gas tightness of this prototype was verified beforehand with the leak test described in Section~\ref{sec:prototypes}.

\begin{figure}[H]
\centering
\includegraphics[width=0.8\columnwidth]{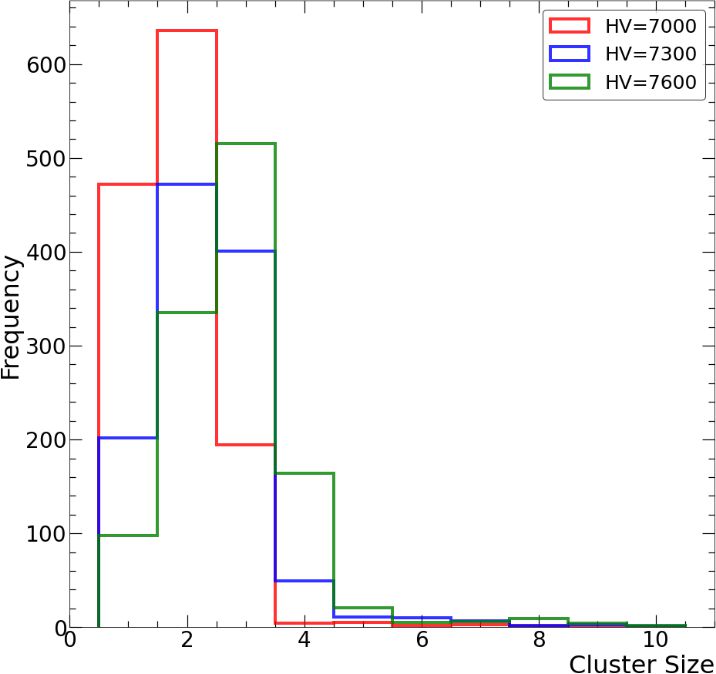}
\caption{Cluster size distributions of prototype C for cosmic-ray muons at applied voltages of 7.0, 7.3 and 7.6~kV~\cite{Samalan2024_thesis}.}
\label{fig:clusterC}
\end{figure}

The cluster size, defined as the number of adjacent strips fired per event, was measured at 7.0, 7.3 and 7.6~kV with 10000 triggers per point (Fig.~\ref{fig:clusterC}). Most events have a cluster size between one and four strips. The distribution shifts slightly towards larger clusters as the voltage increases, with the most probable value moving from two strips at 7.0~kV to three strips at 7.6~kV. 

\begin{figure}[H]
\centering
\includegraphics[width=0.8\columnwidth]{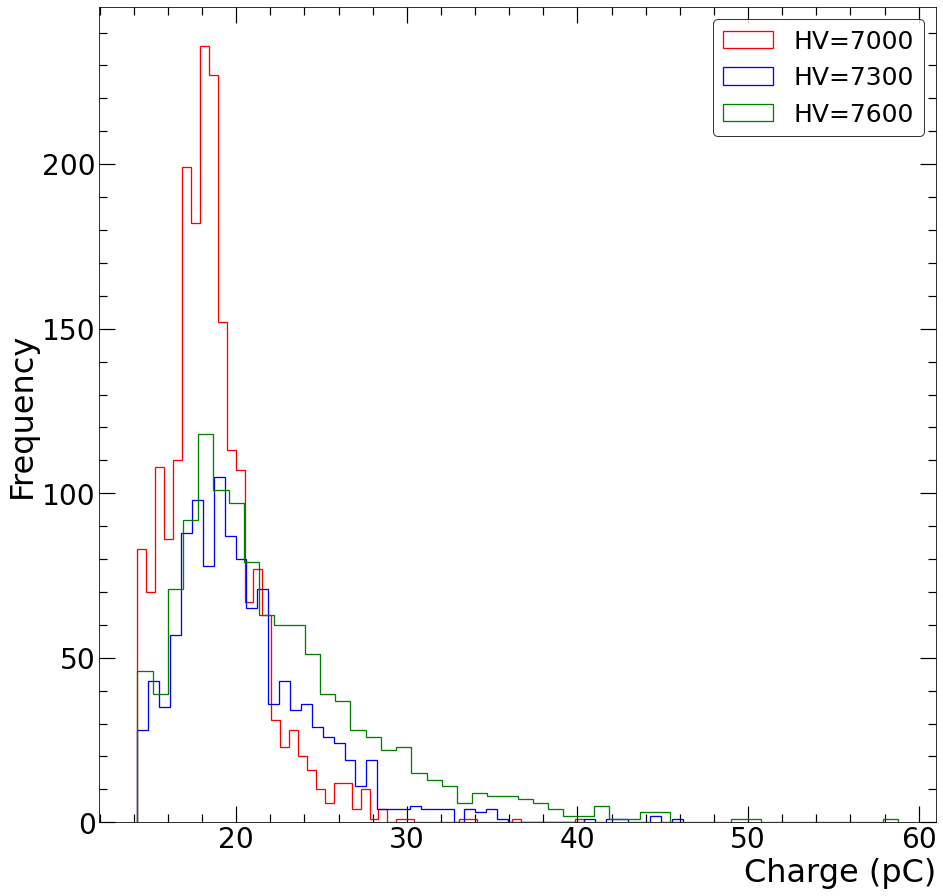}
\caption{Pulse charge distributions at applied voltages of 7.0, 7.3 and 7.6~kV ~\cite{Samalan2024_thesis}.}
\label{fig:chargeC}
\end{figure}

The QDC also gives the charge induced on each strip, with one QDC count corresponding to 0.098~pC. Figure~\ref{fig:chargeC} shows the pulse charge distributions at the same three voltages. At higher voltage, the high-charge tail of the distribution grows. The cluster charge is the sum of the charges of the strips belonging to a cluster. Its mean value rises from about 21~pC at an applied voltage of 6.4~kV to about 33~pC at 7.0~kV and about 58~pC at 7.6~kV ~\cite{Samalan2024_thesis}.

\begin{figure}[H]
\centering
\includegraphics[width=\columnwidth]{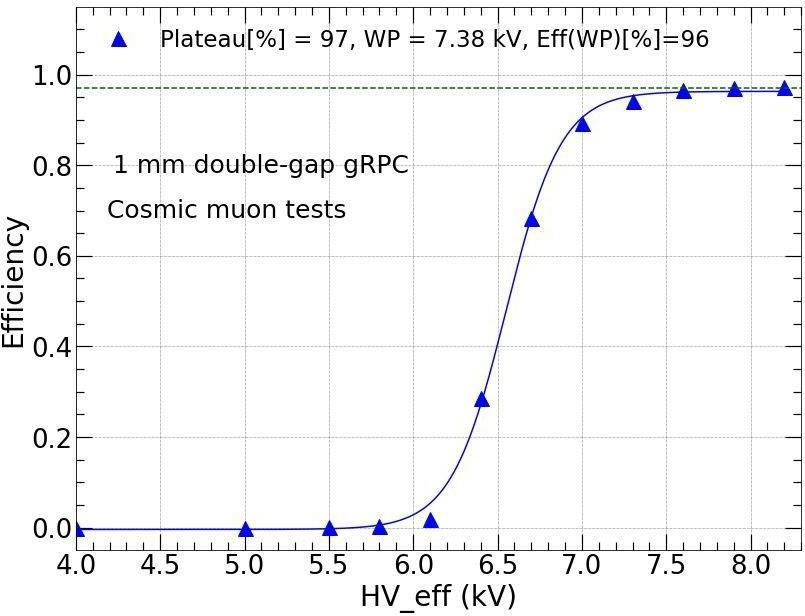}
\caption{Cosmic-ray muon detection efficiency of the 1~mm double-gap prototype C as a function of the effective high voltage ~\cite{Samalan2024_thesis}.}
\label{fig:effC}
\end{figure}

The detection efficiency was measured with 1000 triggers per voltage point (Fig.~\ref{fig:effC}). The efficiency starts to rise above 6~kV and reaches a plateau of about 97\% above 7.5~kV. At the working point of 7.38~kV the efficiency is 96\%, which shows that the double-gap layout performs well.

\section{Muography-like Absorption Test}\label{sec:absorption}

\begin{figure}[H]
\centering
\includegraphics[width=\columnwidth]{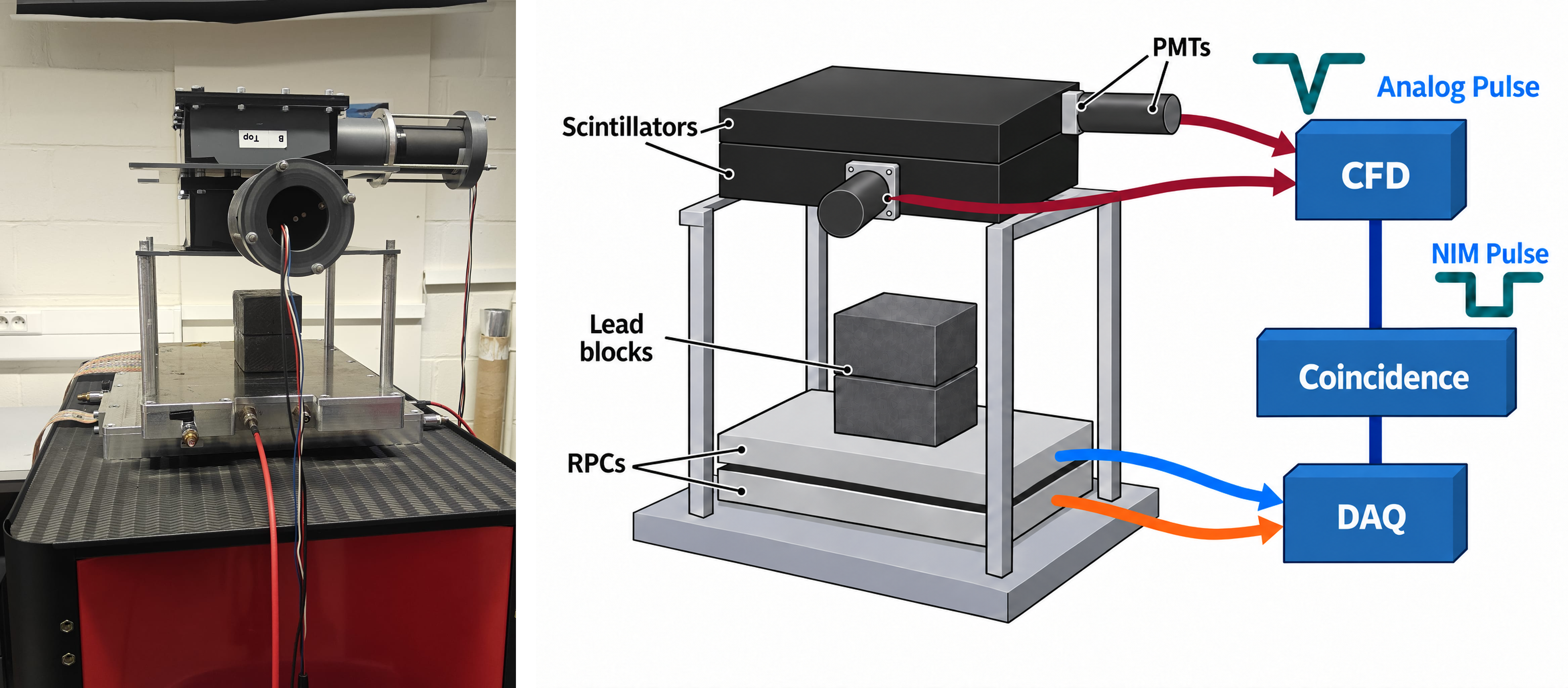}
\caption{Left: experimental setup of the muon absorption studies with prototype A. Right: schematic of the RPC stack with orthogonal $X$ and $Y$ strip readout planes traversed by a cosmic-ray muon.}
\label{fig:absSetup}
\end{figure}

To assess the imaging capability of the system, a simple muography-like absorption measurement was performed in the laboratory with prototype A~\cite{Ikram2025}. The setup consists of two plastic scintillators read out by PMTs, which provide the trigger, and two layers of four RPCs with $X$--$Y$ readout placed below them (Fig.~\ref{fig:absSetup}). Two lead blocks of 10$\times$7.5$\times$4.5~cm$^3$ were placed as absorbers between the scintillators and the RPCs. Data were measured in two configurations: without absorber (``free sky''), which serves as reference, and with the lead blocks in place. The strip occupancy distributions measured under free-sky conditions and with the lead absorber were normalized to a common exposure of 24 hours. The 1D occupancy plots present the normalized muon counts as a function of strip fired for each detector layer.

\begin{figure}[H]
\centering
\includegraphics[width=.9\columnwidth]{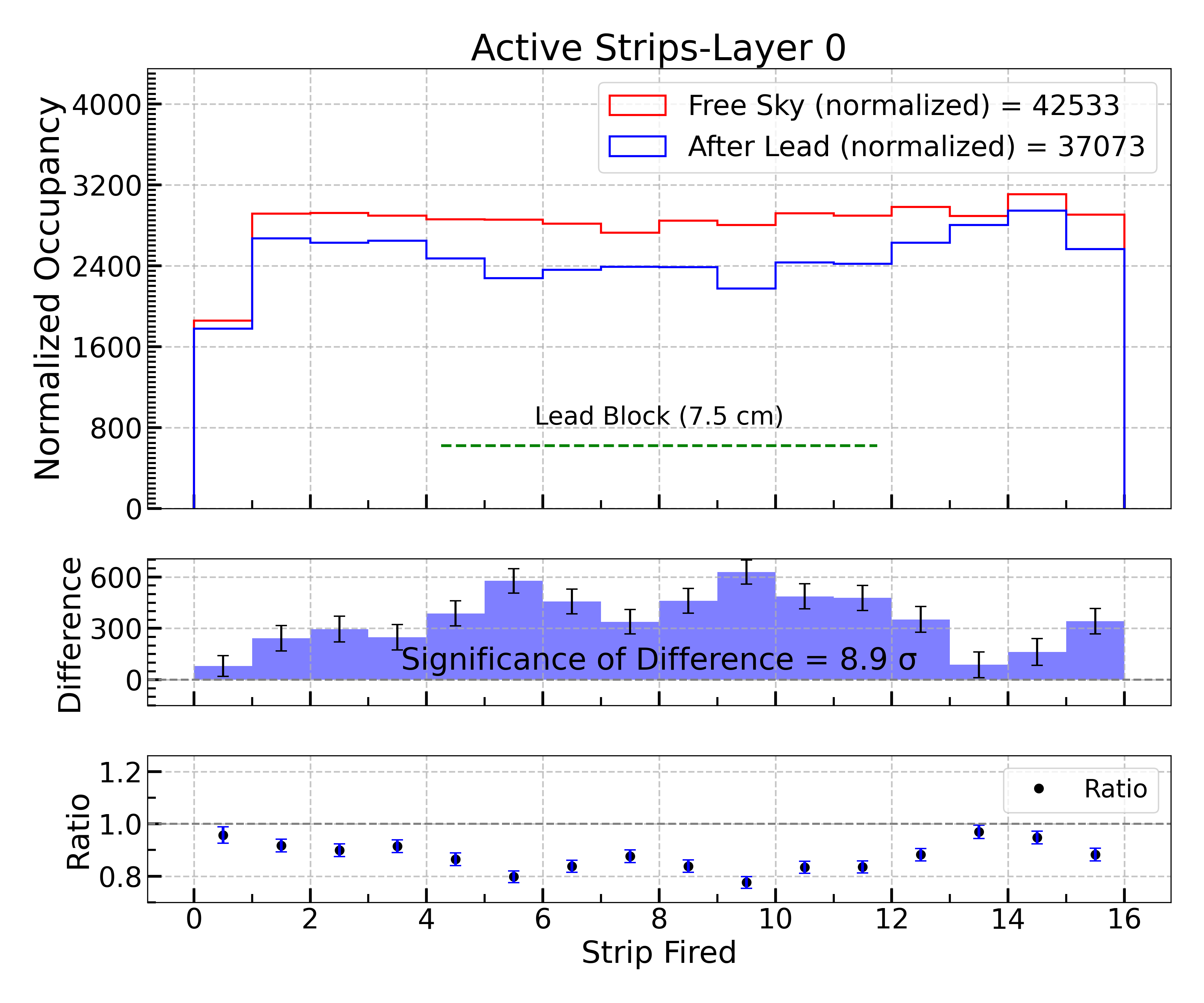}
\caption{Normalized strip occupancy of the top RPC (layer 0) for the free-sky (red) and lead-block (blue) configurations (top), their difference (middle) and their ratio (bottom). The dashed green line indicates the position and extent (7.5~cm) of the lead blocks.}
\label{fig:absTop}
\end{figure}

Figure~\ref{fig:absTop} shows the normalized strip occupancy of the top RPC (layer 0) for the free-sky and lead-block configurations, together with their difference and ratio. A clear reduction of the normalized occupancy is observed in the region covered by the lead blocks, where the occupancy ratio decreases to about 0.8. A smaller reduction is also visible in the neighbouring strips, which is attributed to inclined muons crossing the absorber.

\begin{figure}[H]
\centering
\includegraphics[width=.9\columnwidth]{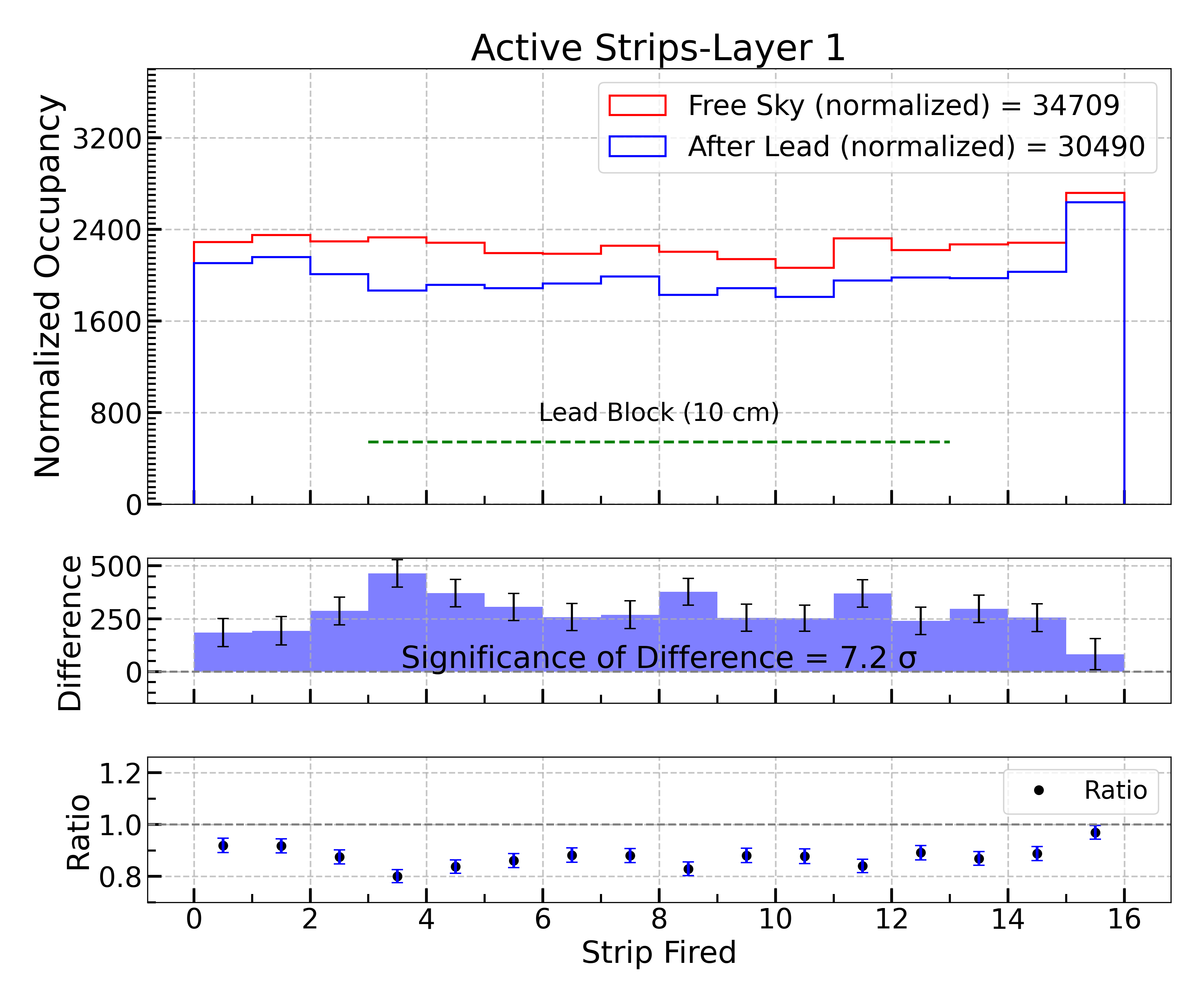}
\caption{Normalized strip occupancy of the bottom RPC (layer 1) for the free-sky (red) and lead-block (blue) configurations (top), their difference (middle) and their ratio (bottom). The dashed green line indicates the position and extent (10~cm) of the lead blocks.}
\label{fig:absBottom}
\end{figure}

The same behaviour is observed in the bottom RPC (layer 1) (Fig.~\ref{fig:absBottom}), where the lead blocks extend over 10~cm in the direction perpendicular to the strips. In both layers, the decrease of the normalized occupancy due to the lead absorber is clearly visible, demonstrating the sensitivity of the system to the presence of high-density material and its potential for absorption muography.

\section{Summary and Outlook}\label{sec:summary}

We present the development of portable muon telescopes based on sealed glass RPCs, targeting muography applications in remote outdoor locations and confined spaces. Three small-area gRPC prototypes with different electrode, gap, and readout configurations have been constructed and characterized. The design aims at a modular, mechanically robust and lightweight detector suitable for operation with low gas flow or in fully sealed mode.

The laboratory measurements with cosmic-ray muons confirm the performance of the three designs. Prototype A, operated in sealed mode, reaches a maximum detection efficiency of about 85\% at the working point of 6.8~kV and maintains a stable performance over more than two years, with periodic gas refills compensating for the gradual efficiency loss. Prototype B, operated in a three-layer telescope, achieves a detection efficiency of at least 95\% above 10~kV, a time resolution of about 1.4~ns and a position resolution of about 3~mm, while the acrylic casing stabilizes the relative humidity inside the chamber. The 1~mm double-gap prototype C reaches an efficiency plateau of about 97\% above 7.5~kV, with an average cluster size of about two strips, and the leak test of its 3D-printed, cable-free gas volume shows no measurable pressure loss at an argon overpressure of 22~mbar over about 65 hours. These results demonstrate that the leak-tight designs and the operation in sealed mode perform as intended. A first absorption measurement with prototype A shows a clear reduction of the normalized strip occupancy in the region shadowed by the lead blocks, confirming the sensitivity of the system to high-density materials.

Future developments will focus on four aspects. First, the readout granularity will be increased by new readout boards, such as the 64-strip boards, by optimizing the board layout through signal multiplexing, and potentially by the MAROC3A-based readout board, which is currently under development. Second, the 3D-printed frame solution will be extended to further reduce the gas volume. Third, the standard gas mixture will be replaced by environmentally friendly RPC gas mixtures that eliminate high global-warming-potential components~\cite{Ahmadi:2025ecogas, Quaglia2024}. Finally, the telescopes will be deployed in outdoor applications, which requires addressing the associated logistical constraints of field operation.

\par\vspace{12pt}
\noindent{\bfseries\large Acknowledgment\par}
The VUB team acknowledges the support from the VUB Research Council via Grant No. OZR4149. The CP3 team's contribution to this work was partially supported by the EU Horizon Europe Innovation Action under grant agreement No. 101233379 (``KINETIKA''). NISER's work was partially supported by the Department of Atomic Energy, India. B. Mohanty acknowledges the JC Bose Fellowship of the Anusandhan National Research Foundation, India, for the support.

\vspace{4pt}
\noindent
\bibliographystyle{JHEP}
\bibliography{references}

\end{document}